\documentclass[preprint]{aastex}

\shorttitle{Close binary stars.~VI}
\shortauthors{Rucinski \& et al.}

\begin{document}

\title{Radial Velocity Studies of Close Binary 
Stars.~VI\footnote{Based on the data obtained at the David Dunlap 
Observatory, University of Toronto.}}

\author{Slavek M. Rucinski, Wenxian Lu\footnote{Current address:
  NASA/GSFC, Bldg.\ 26, 8800 Greenbelt Road, Greenbelt, 
MD 20771}, \  Christopher C.\ Capobianco, \\
Stefan W. Mochnacki,
R.\ Melvin Blake, J. R. Thomson}
\affil{David Dunlap Observatory, University of Toronto\\
P.O.Box 360, Richmond Hill, Ontario, Canada L4C 4Y6}
\email{rucinski,lu,capobian,mochnacki,blake,jthomson@astro.utoronto.ca}

\and

\author{Waldemar Og{\l}oza, Greg Stachowski}
\affil{Mt. Suhora Observatory of the Pedagogical University\\
ul.~Podchora\.{z}ych 2, 30-084 Cracow, Poland}
\email{ogloza,stachowski@ap.krakow.pl}


\begin{abstract}

Radial-velocity measurements and sine-curve fits to the orbital 
velocity variations are presented for the sixth set of ten close binary 
systems: SV~Cam, EE~Cet, KR~Com, V410~Cyg, GM~Dra, V972~Her, ET~Leo,
FS~Leo, V2388~Oph, II~UMa. 
All systems except FS~Leo are double-lined spectroscopic binaries.  
The type of FS~Leo is unknown while
SV~Cam is a close, detached binary; all remaining systems are
contact binaries. Eight binaries (all except SV~Cam and V401~Cyg)
are the recent photometric discoveries of the Hipparcos satellite 
project. Five systems, EE~Cet, KR~Com, V401~Cyg, V2388~Oph, II~UMa,
are members of visual/spectroscopic triple systems. We were able
to observe EE~Cet separate from its companion, but in the 
remaining four triple systems we could separate the spectral 
components only through the use of the broadening-function approach.
Several of the studied systems are prime candidates for combined 
light and radial-velocity synthesis solutions.
\end{abstract}


\section{INTRODUCTION}
\label{sec1}

This paper is a continuation in a series of papers of
radial-velocity studies of close binary stars:
\citet[Paper I]{ddo1}, \citet[Paper II]{ddo2} and
\citet[Paper III]{ddo3}, \citet[Paper IV]{ddo4}, 
\citet[Paper V]{ddo5}.
The main goals and motivations are described in these papers.
The companion paper of \citet[Paper VII]{ddo7} describes the technical
details and methods of data reductions as well as the resulting
measurement uncertainties, based on data in the six previous papers.

This paper is structured in the same way as Papers~I -- V in that 
most of the data for the observed binaries are in 
two tables consisting of the radial-velocity measurements 
(Table~\ref{tab1}) and their sine-curve solutions (Table~\ref{tab2}).
Section~\ref{sec2} of the paper contains brief summaries of 
previous studies for individual systems. Figures~\ref{fig1} --
\ref{fig3} show the data and the radial velocity solutions.

The observations reported in this paper have been collected between
October 1997 and June 2001; the ranges of dates for individual 
systems can be found in Table~\ref{tab1}. 
All systems discussed in this paper, except SV~Cen, 
have been observed for radial-velocity variations for the first time. 
We have derived the radial velocities in the same way as described 
in previous papers. See Paper~IV and Paper~VII for discussion of the
broadening-function approach used in the derivation of
the radial-velocity orbit parameters:
the amplitudes, $K_i$, the center-of-mass
velocity, $V_0$, and the time-of-eclipse epoch, $T_0$.

The data in Table~\ref{tab2} are organized in the same manner as 
in previous papers. In addition to the parameters of spectroscopic orbits,
the table provides information about the relation between 
the spectroscopically observed epoch of the primary-eclipse T$_0$ 
and the recent photometric determinations in the form of the (O--C) 
deviations for the number of elapsed periods E. It also contains
our new spectral classifications of the program objects. 

For further technical details and conventions used in the paper, please
refer to Papers I -- V and VII of this series.

\placetable{tab1}
\placetable{tab2}

\placefigure{fig1}
\placefigure{fig2}
\placefigure{fig3}

\section{RESULTS FOR INDIVIDUAL SYSTEMS}
\label{sec2}

\subsection{SV Cam}

SV Cam is a short-period, detached binary system. It has been the subject
of numerous photometric and four spectroscopic  
studies. In two first spectroscopic investigations, 
\citet{hil53} and \citet{rai91}, the system was
observed as a single-lined spectroscopic binary (SB1). The secondary
component was subsequently detected by \citet{poj98} who 
used the spectral resolution three times lower than ours. \citet{oez01}
observed the system at a relatively high spectral resolution, but chose
to adopt the spectroscopic results of \citet{poj98}, concentrating on
the details of line-profile changes, in relation to the known, high
activity of this short-period RS~CVn-type system.

Our double-lined (SB2) 
spectroscopic orbit is very well defined, with small errors
of the orbital parameters. We note that determinations of
the radial velocity semi-amplitude of the primary component, $K_1$, agree
well in the four existing solutions. Starting with \citet{hil53}
(as reanalyzed by \citet{rai91}), \citet{rai91}, \citet{poj98} 
and the current, the results have been (in km~s$^{-1}$) : 
$121.7 \pm 1.9$, $122.3 \pm 1.5$,
$118.5 \pm 2.0$, and $121.86 \pm 0.76$. 
The results of \citet{poj98} differ the most, but are still  
within the errors of the solutions. 
The center of mass velocity, $V_0$, seems to show a secular progression,
although the data of \citet{poj98} deviate from the trend. In the same
order as before: 
$-16.2 \pm 1.4$, $-11.2 \pm 1.2$, $-13.7 \pm 1.5$ and
$-9.13 \pm 0.78$ km~s$^{-1}$. Noting that the observations were
made in 1947, 1988, 1993 and 1997, the trend may be related to the
motion about the third body with the period of about 65 -- 75 years
and the semi-amplitude of about 2 km~s$^{-1}$ \citep{rai91}.

The results of \citet{poj98}, although the first to reveal the
motion of the secondary component, show deviations described
above and, even more importantly, differ rather substantially from our
results in the value of $K_2$: $211.5 \pm 5.5$ versus
our $190.17 \pm 1.73$ km~s$^{-1}$. Possibly, the discrepancies 
are due to the rather complex reduction scheme of the previous 
study which involved successive removal of the two
spectral signatures from individual spectra. This was
entirely unnecessary in our case because the broadening functions 
that we analyzed were very well defined and radial velocities of both 
components could be measured with great ease. 
Figure~\ref{fig4} shows one of the broadening functions for SV~Cam,
in comparison with other systems analyzed in this paper. On the basis
of the widths of the individual signatures in the
broadening functions, and taking into consideration the instrumental
broadening, we estimate the apparent rotation velocity of the
components, $V_1 \sin i = 122 \pm 10$ km~s$^{-1}$ and 
$V_2 \sin i = 85 \pm 8$ km~s$^{-1}$. \citet{pat94} found that
the orbit of SV Cam is oriented exactly edge on, $i \simeq 90^\circ$,
so that these are our estimates of the equatorial rotational
velocities. We note that \citet{poj98} estimated 
$V_1 \sin i = 105$ km~s$^{-1}$ while a solar-size
star in the orbital synchronism would have 
$V \sin i \simeq 90$ km~s$^{-1}$.

The total mass, $(M_1+M_2) \sin^3i = 1.871 \pm 0.045\,M_\odot$,
and the mass ratio, $q=0.641 \pm 0.007$, lead to the masses,
1.14 and 0.73 $M_\odot$, which are somewhat large for the 
combined spectral
type that we estimated as G2V, but would agree with the estimate of
\citet{poj98} of F8V. The estimated value of $V_1 \sin i$ would
also suggest a primary larger than the Sun.
The mean color index from the Tycho-2 Catalog \citep[TYC2]{tyc2},
$B-V=0.62$, is in agreement with our spectral type.
We should note that our low-dispersion classification as well
as the TYC2 color index relate to the combined photometric properties 
of the system. 

The ephemeris that we used was that of the photometric study of
\citet{pri01} which covered the time range actually slightly after our
observations and gave a perfect agreement for the time of eclipses.
Since the published $T_0$ was already shifted from
the actual time of observations, we 
recalculated the published moment to our $T_0$ so that no whole
cycles appear in the phase count in Table~\ref{tab2}.
We also used the orbital period from the same study simplifying it to the
six decimal places, 0.593073 days, which was entirely sufficient for the
duration of our observations. We note that the orbital period cited
in the Hipparcos Catalogue \citep[HIP]{hip} 
was slightly different, 0.593075 days,
while \citet{poj98} used 0.593071 days.

With a good Hipparcos parallax, $p=11.77 \pm 1.07$ mas (HIP), 
and well determined tangential motions, 
$\mu_\alpha \cos \delta = 41.5 \pm 1.1$ mas/y and 
$\mu_\delta = -150.9 \pm 1.1$ mas/y (TYC2), the system of SV Cam 
appears to be one of the best currently characterized close binary
systems on the lower main sequence. 
Of note is the relatively large spatial motion of 64 km~s$^{-1}$,
mostly the result of the large tangential motion; the systemic
radial velocity is moderate, $V_0=-9.13 \pm 0.78$ km~s$^{-1}$.

\subsection{EE Cet}

Variability of the combined light of the visual binary
ADS 2163 was discovered
by the Hipparcos satellite mission \citep{hip}. The light curve was
relatively sparsely covered, showing variation of about 0.23 magnitude.
On the basis of an observation that the southern component rotates too 
rapidly for the technique used by \citet{nor97} 
to measure its radial velocity, this component
was our prime candidate for being the source of the variability
observed by Hipparcos. We could observe this star
without contamination by the northern companion which is 
separated by 5.6 arcsec. We found that the
southern component is a contact binary system and we assume that
it should carry the variable-star name EE~Cet. It appears to be 
a contact system of the W-type, i.e.\ with the 
less massive component eclipsed during the primary eclipse; 
however, it is the more massive, slightly 
cooler component which gives a better defined signature in the 
broadening function, probably because of its larger radiating
area. In phasing our observations, we used the HIP prediction for
times of eclipses.

We caution that there exists certain confusion in the literature as to
which component of ADS~2163 should be called A and which B.
On the average, the southern component, which we identify here with
EE~Cet, is the fainter of the two and this agrees with the naming of 
the stars in the HIP catalog, where the visual binary 
appears under CCDM J02499+0856. However, in the ADS catalog 
the names are actually reversed and in SIMBAD it is the
northern, brighter component which carries the name of EE~Cet.
\citet{lam01} published photometric data for one epoch and 
give $V(A)=9.47$ and $V(B)=9.83$ identifying the southern component
as fainter.
To complicate things even further, we found that the northern component 
of ADS 2163 is also a close binary system showing radial
velocity variations; this component may very well be
a variable star. However, currently we have insufficient 
radial-velocity data to analyze this binary system; 
we hope to be able to provide such data in one of the subsequent 
papers of this series. We only note that \citet{nor97} gave its
radial velocity, $V_0 = +2.26 \pm 0.04$ km~s$^{-1}$, 
which is similar to the center-of-mass velocity of EE~Cet, 
$V_0 = +1.60 \pm 0.93$ km~s$^{-1}$. 

The radial velocity solution for EE~Cet is very well defined. 
However, because of the lack of reliable Hipparcos parallax 
(it is actually listed as negative), we cannot derive 
the absolute magnitude of the binary. Once the two components of the
visual binary are photometrically separated and parallaxes for
both components determined, this quadruple system may turn out as
one of the most interesting among bright stars of the sky.

\subsection{KR Com}

KR Com is another Hipparcos satellite photometric discovery. 
In phasing our observations, we used the HIP prediction for
times of minima.
The shallow eclipses are of almost identical depth of 
only about 0.06 mag.\ so there is no obvious reason for 
the EB classification which has been used in the HIP Catalog;
the system is a contact binary and would normally be designated as EW. 

The star had been known 
as a very close visual system of very small mean separation of 
only 0.12 arcsec. In the HIP catalog it appears as CCDM J13203+1746, with
the magnitude difference of components of 0.60 magnitude; it is 
also known as ADS 8863. The visual binary was a subject of numerous
speckle interferometry investigations. While these observations
are of lesser relevance in the context of this paper, we note that
the close binary shows rather large radial-velocity variations, so that 
its small photometric variation may be partly due to ``dilution'' of the
close-binary variability signal in the combined light of the
visual system. Our broadening function indicates that the brighter
component is the contact system while the fainter component is
a slowly-rotating star, as can be seen in Figure~\ref{fig4}. 
The brightness of the visual companion is large, 
$L_3/(L_1+L_2)=0.56 \pm 0.04$ (at light maxima of the close binary), 
which confirms the previous magnitude-difference estimates.

Our radial velocity orbit is very well defined, mostly due to the
brightness of the system at $V_{max}=7.14$. We could very
well isolate the signature of the third, slowly-rotating star,
as one can see in Figure~\ref{fig4}. Allowing for the contribution of the
visual companion, and with the Hipparcos parallax of 
$p=13.07 \pm 0.87$ mas, one obtains $M_V^{tot}=2.72 \pm 0.15$
and thus $M_V^{EW} = 3.20$, while the RD97 calibration predicts 
$M_V^{EW}(cal)=3.42$. The latter is obtained using the 
color TYC2 index, $(B-V)=0.52$, which suggests a spectral type of about
F8; our direct classification is G0IV, although the spectral type
and the luminosity class relate to the combined properties
of the triple system.  The average radial velocity of
the third component, $\overline{V_3} = -3.59 \pm 0.12$ km~s$^{-1}$
(the error of a single observation is 
0.93 km~s$^{-1}$), is significantly different from 
the center-of-mass velocity of the binary, 
$V_0=-7.86 \pm 0.38$ km~s$^{-1}$.

KR~Com has one of the smallest mass ratios known among contact
binaries, $q=0.091 \pm 0.002$. The total minimum mass, 
$(M_1+M_2) \sin^3i = 0.517 \pm 0.008\, M_\odot$,
indicates that the orbit is rather
strongly inclined to the line of sight, which is another reason for the
small photometric amplitude. As is normally observed, the small mass
ratio is associated with the A-type characteristics of the contact system,
i.e.\ the deeper eclipse corresponds to the more massive star
being eclipsed by the less massive companion.

\subsection{V401 Cyg}

Together with SV~Cam, this binary is one of two in this paper which 
had not been discovered by the Hipparcos mission. In fact it was
discovered by \citet{hof29} and since then was sporadically observed, 
mostly for eclipse timing. Photometrically, the binary 
appears to be a rather typical contact system with relatively large light
variation of about 0.55 magnitude. \citet{her93} pointed out that the
orbital period of the system is lengthening. \citet{wol00} conducted
a light-curve analysis and found the best-fitting, photometric mass ratio, 
$q_{ph} \simeq 0.3$. To some extent to our surprise (because we see so many
cases of $q_{ph} \ne q_{sp}$), we found $q=0.290 \pm 0.011$.

Our radial velocity orbit was based on the assumed period of 0.582714 days,
as in the HIP Catalog. For the initial time of the primary minimum, we used
the data in \citet{nel01} obtained during the span of our
observations. The agreement is very good (see Table~\ref{tab2}) 
given the somewhat
larger error of our $T_0$ for this star, when compared with 
other binaries. The main reason for the larger error
was the relative faintness of this binary ($V_{max} \simeq 10.5$) and the
fact that we discovered a third ``light'' in the system. The 
spectral signature of the third star (which may, but does not have to
be physically associated with the binary)  
can be seen in the broadening function in Figure~\ref{fig4} 
as a small feature projecting onto the prominent signature
of the primary component. Similar noise fluctuations are common in broadening
functions of faint stars, but this one is always present, at the
same radial velocity at all orbital phases.
By integrating the amount of light in this additional
feature, we could estimate that, at the light maximum of the binary,
$L_3/(L_1+L_2) = 0.03 \pm 0.01$. Because the third-light contribution is
so small, we neglected it in our measurement. However, it may have
slightly affected the
amplitude of the primary component, $K_1$, due to the fact that (1) the
radial velocities of primary component are, by necessity, always close
to $V_0$ and (2) the third component appears to
have a very small radial velocity relative to the binary system.

\subsection{GM Dra}

GM Dra was discovered photometrically by the Hipparcos satellite as
a 9th magnitude contact binary with a fairly large variation 
of about 0.27 mag. The system was observed photometrically by
\citet{cic01}, one year after our observations. The authors
gave a recent moment of the light minimum and 
improved the value of the orbital period. 

Our radial velocity orbit is quite well defined (Figure~2). 
GM~Dra is a rather uncomplicated contact system of the W-type, 
i.e.\ with the smaller star eclipsed
during the deeper minimum (although the difference in the depth
of the eclipses is very small). The spectral type F5V and the 
color index $B-V=0.48$ agree. The absolute magnitude derived from the HIP
parallax, $p=10.16 \pm 0.88$ mas and $V_{max}=8.65$ is $M_V=3.68 \pm 0.20$,
which is in perfect agreement with the calibration of \citet[RD97]{RD97},
$M_V (cal) = 3.66$. 

\subsection{V972 Her}

V972 Her is another Hipparcos photometric discovery. The light variation
is small, about 0.08 mag.\ indicating a small orbital inclination or
``dilution'' of light in a triple system. The HIP light curve is very
well defined and suggests a perfect contact system (EW). It is not clear
why the system was given there a classification EB which would suggest
unequally deep minima.

Our radial velocity orbit is very well defined, mostly thanks to the
large brightness of the system, $V_{max}=6.62$. With the relatively
large HIP parallax (the largest in this group of binaries),
$p=16.25 \pm 0.61$ mas, and thus $M_V=2.67 \pm 0.09$,
the system is the nearest contact binary in this group 
of ten binary systems. 
The RD97 calibration with $B-V=0.39$ (which agrees with our 
spectral type F4V) gives $M_V (cal)=2.87$, which is in marginal agreement
with the directly determined $M_V$. The very small value of
$(M_1+M_2) \sin^3i=0.276 \pm 0.006 \,M_\odot$ confirms
the expectation of a very small orbital inclination angle. 

Our initial $T_0$ was based on the time determined by \citet{kes00}
which was obtained in the middle of our spectroscopic run, but
gives a rather
large O--C = +0.0224 days. We have no explanation for this discrepancy
because our $T_0$ is nominally accurate to $0.0016$ days. 
The binary is a contact 
system of the W-type with the less massive component eclipsed at the
deeper minimum; however, the difference in depth is very small.

\subsection{ET Leo}

The photometric variability of the star was discovered by the
Hipparcos mission. The period assigned there was equal to one half
of the true orbital period. Instead of the exact double of
the HIP period, we used the value given by \citet{gom99}
of 0.346503 days. Their initial epoch $T_0$ (obtained three years before
our observations) and the new period predict the moment
of the deeper eclipse which agrees well with our determination
(see Table 2).

Our spectral type, G8V, appears to be late relative to the TYC2 color
index $(B-V)=0.61$. The parallax is large, $p=13.90 \pm 1.44$
mas, but apparent magnitude is only moderate, $V_{max}=9.48$,  
indicating that the system is
intrinsically rather faint: $M_V=5.20 \pm 0.24$. 
The RD97 calibration, utilizing the $(B-V)$ index as above, 
predicts $M_V (cal)=4.01$; an agreement
in $M_V$ would require a color index as large as
$(B-V) \simeq 1.0$. This discrepancy between the
spectral type and the color index remains to be explained.

ET Leo must be observed at a relatively low orbital inclination angle. 
This would explain the small photometric variability of
only 0.06 mag.\ as well as the small radial-velocity amplitudes
leading to $(M_1+M_2) \sin^3i=0.450 \pm 0.012\, M_\odot$. 

\subsection{FS Leo}

The photometric variability of FS~Leo was discovered by the Hipparcos
mission. In spite of our efforts, we have not been able to find 
any spectral signatures of the secondary component in the system, 
so that this is a single-lined binary (SB1); it is the only 
such system in this group of ten binary systems.
The radial velocity of the only visible component is very well
defined. For the initial phasing of our observations, we used the
original HIP prediction which gave a time
deviation O--C = $-0.035$ days.

Our spectral type is F2V and the photometric data derived from
the HIP and TYC databases give $V_{max}=8.94$ and $(B-V)=0.26$,
the latter suggesting a slightly earlier spectral type around F0.
The HIP parallax, $p=6.12 \pm 1.41$ leads to $M_V=2.87 \pm 0.51$.
At this point we are unable to classify the system, but it is
unlikely that this is a contact system. Of interest is the
large spatial velocity of the system of 70 km~s$^{-1}$
which results mostly from the large tangential motion: $71.3 \pm 1.1$
and $-38.7 \pm 1.1$ mas/year in both coordinates (TYC2).

\subsection{V2388 Oph}

This bright, 6th magnitude, very close
visual binary (Fin 381, CCDM J17542+1108)  
has been the subject of many speckle interferometry investigations.
The study of \citet{soed99} resulted in the following set of
the parameters: the separation, $a=0.088$ arsec, the magnitude difference
between components, $\Delta m = 1.80$ mag, the mean HIP magnitude for the
primary $H_p=6.45$, the orbital period, $P=8.9$ years. $\Delta m$
derived in this study was much larger than in several previous speckle
interferometry investigations which usually suggested a number around
0.2 mag. We note that our entirely independent derivation based on
the broadening functions gives $L_3/(L_1+L2) = 0.20 \pm 0.02$, hence
$\Delta m = 1.75 \pm 0.02$ mag., confirming the result of \citet{soed99}.

\citet{rod98} discovered variability of the star, apparently
independently of the HIP discovery. They classified it as a
W~UMa-type system with a relatively long orbital period of 0.802 days.
Newer photometry of \citet{yak00} (used for $T_0$ in Table~\ref{tab2})
confirmed that the HIP times-of-minima prediction; in fact, the HIP
ephemeris gives a slightly smaller O--C for the time of the primary
eclipse. The $ubvy$ photometric data agree with the previous
spectral type of F5Vn assuming no interstellar reddening. 
We would tend to give the star a slightly earlier spectral type, F3V. 
The HIP light curve is very well defined, and has a shape typical for
a contact system, with eclipses 0.30 and 0.25 mag.\ deep.
The relatively large HIP parallax, $p=14.72 \pm 0.81$,
and $V_{max}=6.13$, gives
$M_V=1.97 \pm 0.13$ for the whole triple system hence 
$M_V^{EW} \simeq 2.17$, while the RD97 calibration 
predicts $M_V (cal)=1.78$ for the W~UMa-type
binary, assuming $(B-V)_0=0.41$ from TYC2. 
The system appears to be one of the
most luminous among currently known contact binaries. This
may explain the imperfect agreement in $M_V$ at the high-luminosity
end where the RD97 calibration becomes highly uncertain.

As for other triple-lined systems, 
we measured the radial velocities of the binary after removing
the third-body signature from the broadening function first
and then by analyzing the remaining contact-binary BF. The radial
velocity orbit of the close binary 
is very well defined (Figure~\ref{fig3}) and describes a 
typical A-type contact system with moderately small mass-ratio,
$q=0.186 \pm 0.002$. The minimum mass,
$(M_1+M_2) \sin^3i=1.93 \pm 0.03\, M_\odot$,
when related to the spectral type of about F3--F5,
suggests a rather large orbital inclination angle, 
$i \simeq 90^\circ$. Thus,
in spite of the necessity of dealing with a complicated BF
(Figure~\ref{fig4}), the radial-velocity parameters for V2388~Oph 
are very well defined. 

The radial velocity measurements for the third component show 
some residual dependence on the phase of the close binary which
may indicate some ``cross-talk'' in the measurements; we illustrate
a similar, but larger effect in the case of the next system,
II UMa. The semi-amplitude of the variations which correlate
with the binary phase 
is about 2.5 km~s$^{-1}$, which leads to a slightly
elevated error per a single-observation of 1.50 km~s$^{-1}$;
for a single sharp-line star we could expect an error
at the level of $1.2-1.3$ km~s$^{-1}$ or less. The
mean radial velocity of the third component, 
$\overline{V_3} = -30.64 \pm 0.20$ km~s$^{-1}$,
is significantly different from the center-of-mass 
velocity for the binary, $V_0=-25.88 \pm 0.52$ km~s$^{-1}$ which
may result from the motion on the 8.9-year orbit.

\placefigure{fig4}

\placefigure{fig5}

\subsection{II UMa}

The photometric variability of the star was discovered by the Hipparcos
satellite. The HIP prediction on the time-of-eclipse
served very well for phasing of our observations.
The star is a known, close visual binary, designated as
ADS~8594 or CCDM J12329+5448,
with the separation of component of 0.87 arcsec and difference in
brightness 1.64 mag.

The broadening functions (see Figure~\ref{fig4})
show a well-defined signature of the third component 
with the relative brightness, $L_3/(L_1+L_2) = 0.17 \pm 0.01$,
implying $\Delta m = 1.92 \pm 0.01$ mag. The triple-star
characteristics of the star were handled by us in the same way as for
the previously described star, V2388 Oph. Again, we see some
unwelcome cross-talk in the radial velocities $V_3$, as shown in
the plot versus the close-binary phase in Figure~\ref{fig5}.
This case is a bit more extreme than for V2388~Oph, with 
the error per single observation of 2.86 km~s$^{-1}$ in place
of the expected $1.2-1.3$ km~s$^{-1}$ or less. 
We do not have a ready explanation for the coupling
between the velocities and why the scatter appears to be
increased only within the first half of the orbit. 
The dependence of $V_3$ on the binary-system phase 
has been observed in some triple stars in our
program, but we usually have no simple explanation for its
occurrence. However, we are not very concerned about its
presence because our goal is the radial velocity orbit of the close
binary system, which is in fact very well defined 
(see Figure~\ref{fig3}).
The average velocity of the third component, 
$\overline{V_3}=-16.02 \pm 0.37$ km~s$^{-1}$, 
can be compared with the center-of-mass velocity 
of the close binary, $V_0 = -8.02 \pm 1.10$ km~s$^{-1}$. 

The binary is a long-period ($P=0.825$ days) contact system 
of the A-type. Our spectral type suggests elevated luminosity,
F5III, but classification of strongly broadened spectra of
contact binaries is not easy, especially when combined
with the spectrum of the third component. However, a high  
luminosity would agree with the relatively
large size of the system implied by the long
orbital period. The system is quite distant, because its HIP parallax is 
$p=5.04 \pm 1.81$ mas. With the observed $V_{max}=8.14$ for the
whole system, $M_V=1.65 \pm 0.8$. Using 
$L_3/(L_1+L2) = 0.17 \pm 0.01$, the absolute magnitude of the
close binary is $M_V^{EW}=1.82$, which agrees rather well with
the RD97 calibration predicting $M_V (cal)=1.70$ for the
assumed TYC2 color index $(B-V)=0.40$.

\section{SUMMARY}

The paper presents radial-velocity data and orbital solutions 
for the sixth group of ten close binary systems 
that we observed at the David Dunlap Observatory. 
Only the detached, short-period system SV~Cam 
was observed spectroscopically before.
All systems except FS~Leo are double-lined (SB2) binaries with 
visible spectral lines of both components and all, with
the additional exception of SV~Cam, are contact binaries. 

Although our selection of our targets is quite unsystematic
and is driven only by the brightness above the limit of
about 11 mag.\ and the shortness
of the orbital period (below one day), 
we continue seeing very interesting objects 
among the bright, photometrically discovered binaries. 
Eight of the systems are new photometric discoveries of the
Hipparcos project. The magnitude-limited nature of the HIP 
survey has led to an emphasis on relatively luminous,
massive, early-type (middle-A to early-F) contact systems. Because
the Hipparcos mission discovered mostly small photometric
amplitude systems, which were overlooked in previous whole-sky
surveys, we tend to observe systems somewhat pre-selected 
in that they either have low orbital 
inclinations, such as V972~Her or ET~Leo, or are members
of triple systems as for KR~Com, V401~Cyg, V2388~Oph and II~UMa. 
In the latter case, while radial velocity
amplitudes are large, the diminished 
photometric variation is due to the ``dilution'' of
the variability signal in the combined light of the triple system. 
In addition to these four triple systems,
we also observed separately a component
in a wider visual binary, EE~Cet; the other component of this
system appears to be radial-velocity variable whose properties
remain to be studied.

\acknowledgements

Support from the Natural Sciences and Engineering Council of Canada
to SMR and SWM is acknowledged with gratitude. 

The research has made use of the SIMBAD database, operated at the CDS, 
Strasbourg, France and accessible through the Canadian 
Astronomy Data Centre, which is operated by the Herzberg Institute of 
Astrophysics, National Research Council of Canada.

We thank the anonymous referee for the very careful and constructive
review of the paper.

\clearpage

\noindent
Captions to figures:

\bigskip

\figcaption[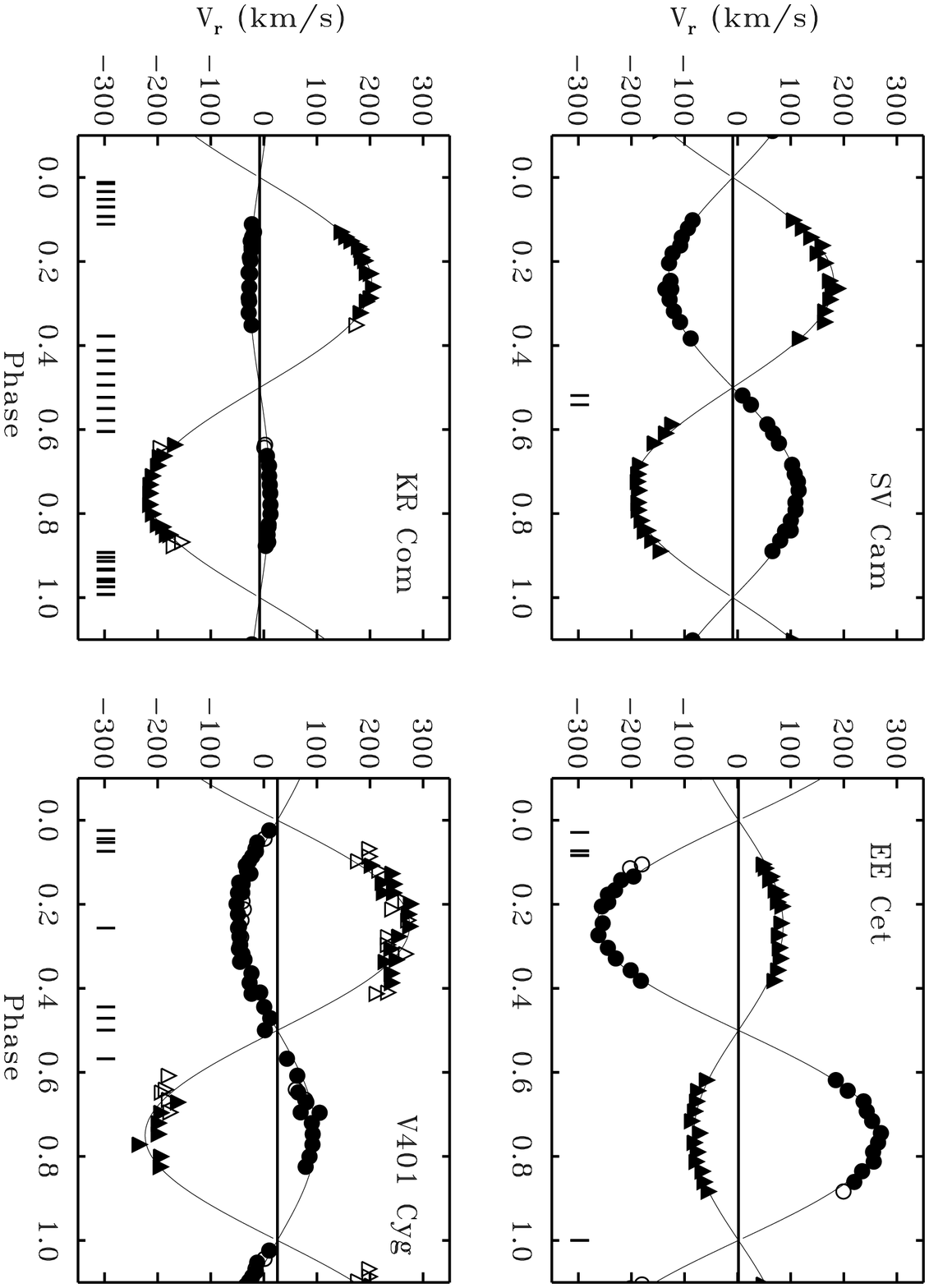] {\label{fig1}
Radial velocities of the systems SV~Cam, EE~Cet, KR~Com and
V401~Cyg are plotted in individual panels versus orbital phases. 
The lines give the respective circular-orbit (sine-curve) 
fits to the radial velocities. SV~Cam is a close, detached binary.
KR~Com and V401~Cyg are A-type contact systems, 
while EE Cet is a W-type contact system. EE~Cet, KR~Com and V401~Cyg
are members of triple systems although the companion of
EE~Cet could be excluded by the spectrograph slit. 
The circles and triangles in this and the next two figures
correspond respectively to components eclipsed at 
the minimum corresponding to $T_0$ 
(as given in Table~\ref{tab2}) or half a period later, 
while open symbols indicate observations contributing 
half-weight data in the solutions. Short marks in the lower 
parts of the panels show phases of available 
observations which were not used in the solutions because of the 
blending of lines. All panels have the same vertical 
ranges, $-350$ to $+350$ km~s$^{-1}$. 
}

\figcaption[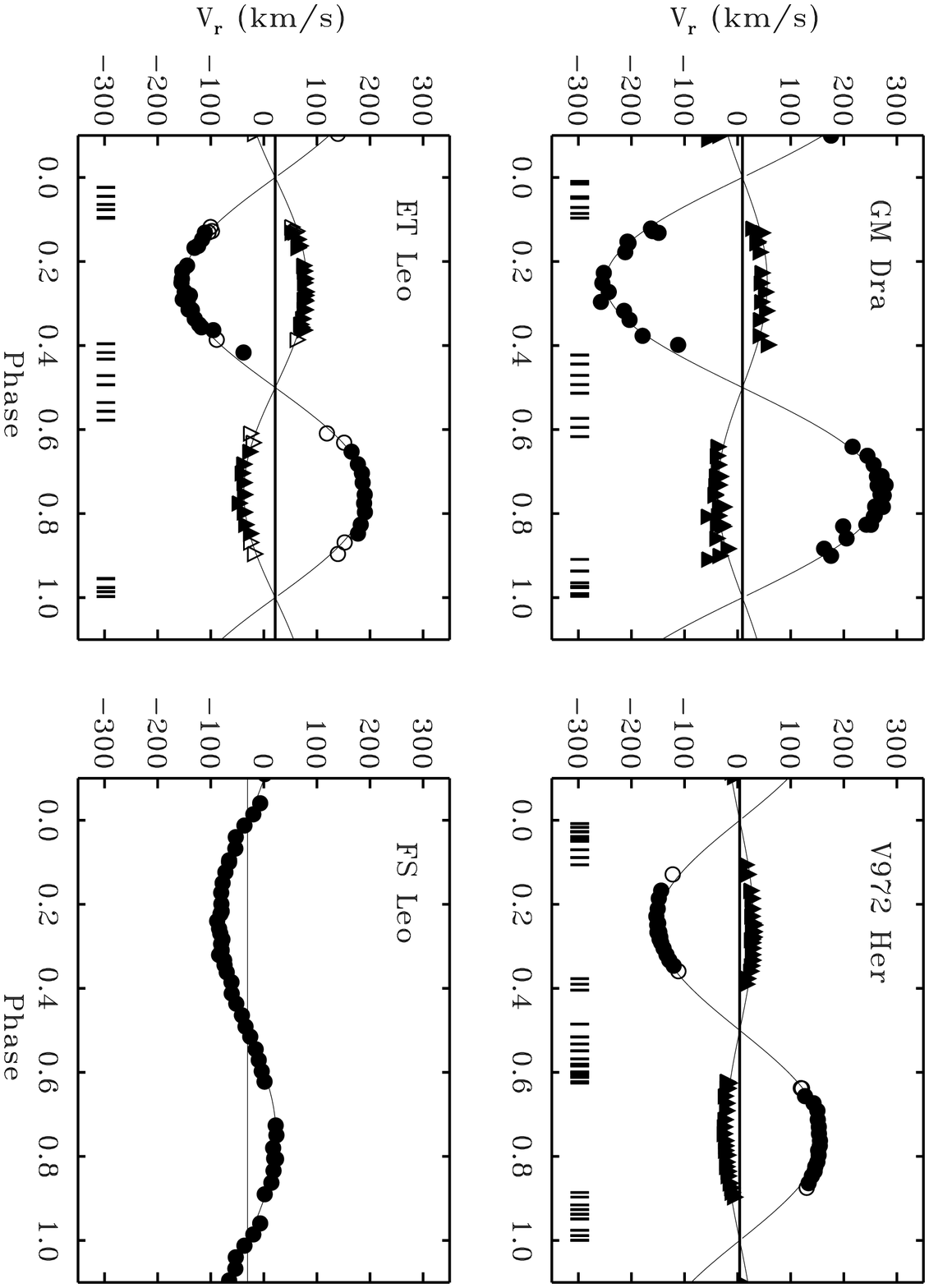] {\label{fig2}
Same as for Figure~1, but with the radial velocity orbits
for the systems GM~Dra, V972~Her, ET~Leo and FS~Leo.
All systems, except FS~Leo (which is a single-line binary 
of unknown type), are W-type contact systems.
}

\figcaption[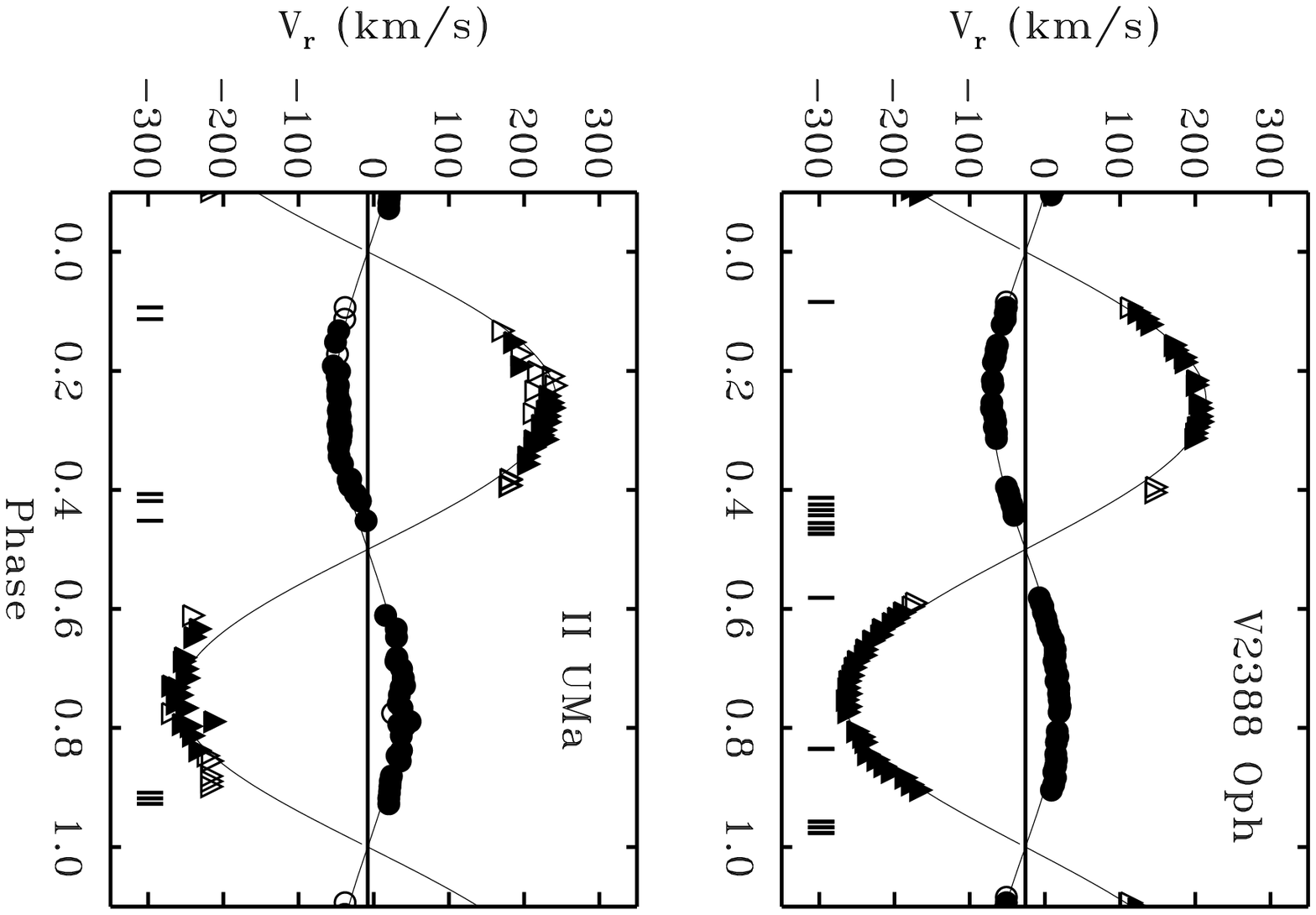] {\label{fig3}
Same as for Figure~1, for the systems V2388~Oph and II~UMa.  
They are quite similar A-type contact systems with relatively
long periods of slightly over 0.8 days, both being members of 
very close triple systems. 
}

\figcaption[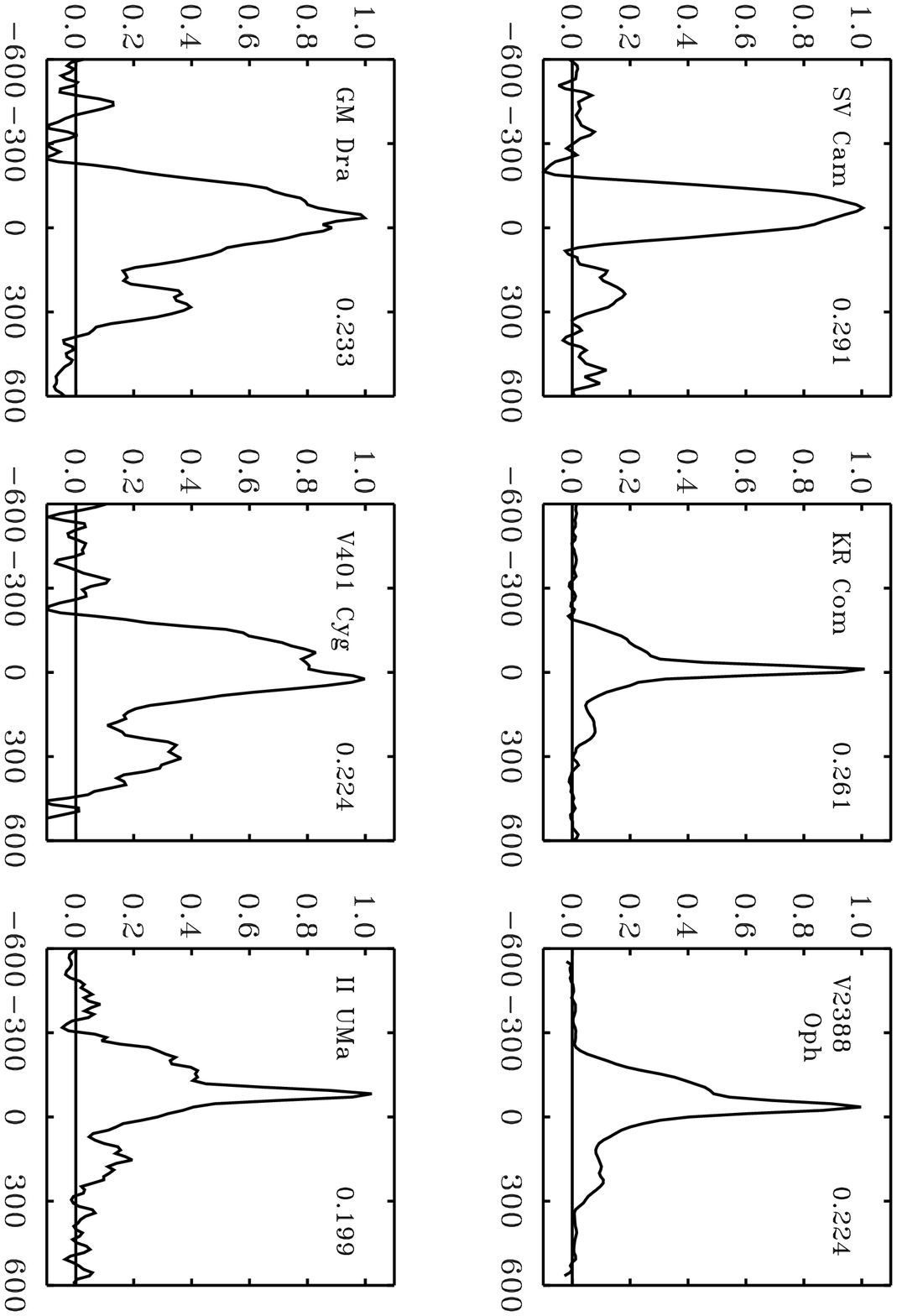] {\label{fig4}
The broadening functions (BF's) for six close binary systems of this
series. The two leftmost panels show the BF's
for rather typical cases of a detached binary SV~Cam 
and the contact binary GM~Dra, both 9 magnitude objects.
The spectroscopic triple systems are shown in the four
center and right panels: The contact binary components
result in broad BF's while the third body produces a narrow
feature between the two broad components.
Note the exceptionally well-defined BF's
for the bright, 6 magnitude systems KR~Com and V2388~Oph. All cases
shown here illustrate orientations of components close to 
the orbital phase 0.25. The small third-body feature in the
BF for V401~Cyg could be taken for a small error fluctuation 
if not for its persistence throughout all the phases.
}

\figcaption[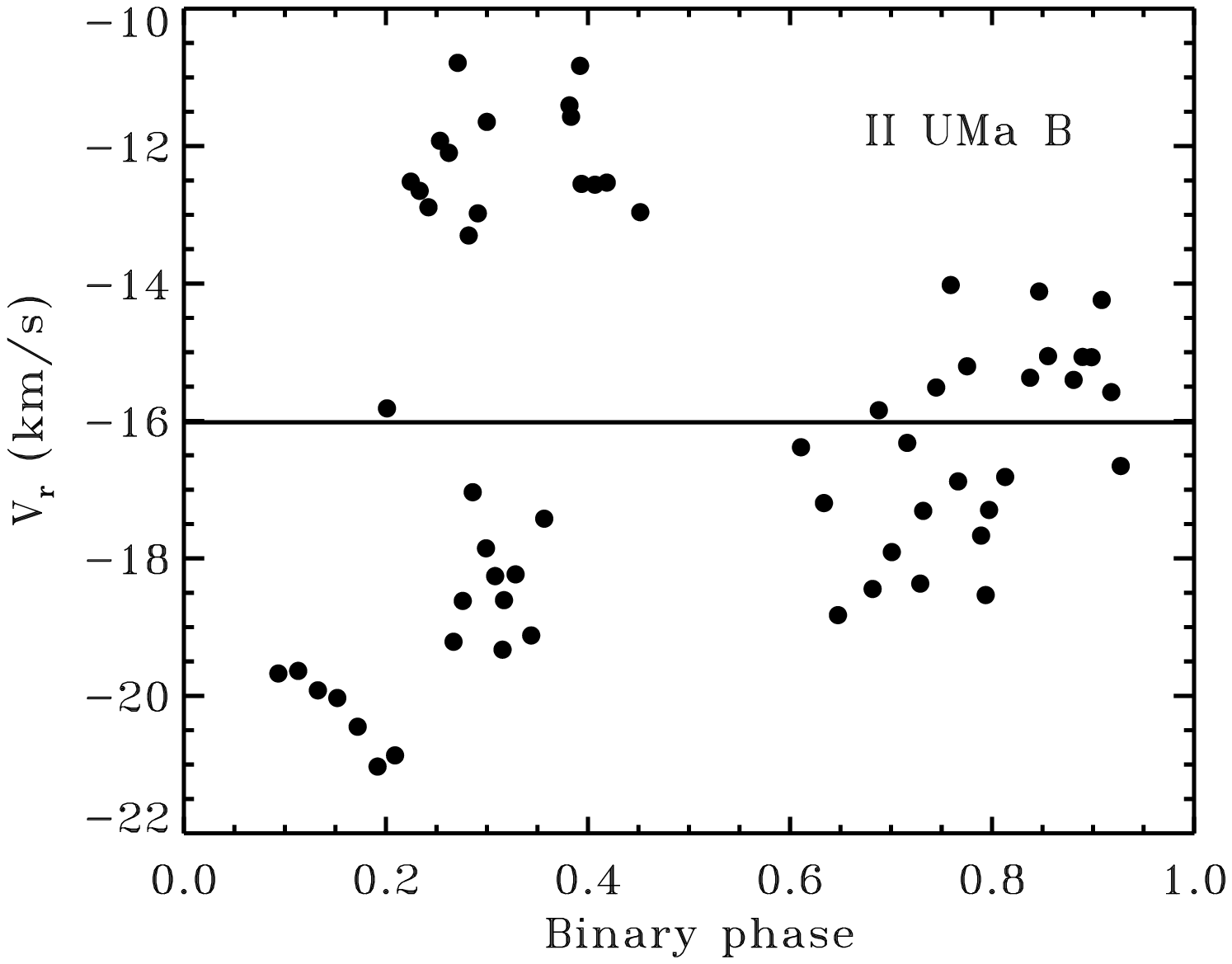] {\label{fig5}
The radial velocities of the companion of II~UMa
plotted versus the orbital phase of the 
close binary. In principle, there should be
no relation between the two quantities, and thus we see here
some sort of a ``cross-talk''. We have no explanation for the
systematic trend in the deviations and, in particular, why
the spread was larger for phases in the first half of the binary
orbit. A similar, but smaller effect has been observed for V2388~Oph.
}

\clearpage                    

\begin{table}                 
\dummytable \label{tab1}      
\end{table}
 
\begin{table}                 
\dummytable \label{tab2}      
\end{table}


\begin{thebibliography}{}

\bibitem[\c Ci\c cek et al.(2001)]{cic01}
    \c Ci\c cek, C., Demircan, O., Erdem, A., \"Ozdemir, S.,
    Bulut, I., Soydugan, E., Soydugan, F. \& Keskin, V.
    2001, Inf. Bull. Var. Stars, 5144
\bibitem[ESA(1997)]{hip}                                       
    European Space Agency. 1997. The Hipparcos and Tycho
    Catalogues (ESA SP-1200)(Noor\-dwijk: ESA) (HIP)
\bibitem[Gomez--Forrellad et al.(1999)]{gom99}   
    Gomez--Forrellad, J. M., Garcia--Melendo, E., Guarro--Flo, J., 
    Nomen--Torres, J. \& Vidal--Saintz, J.       
    1999, Inf.\ Bull.\ Variable Stars, 4702
\bibitem[Herzeg(1993)]{her93}                                  
    Herzeg, T. J.
    1993, \pasp, 105, 911
\bibitem[Hiltner(1953)]{hil53}                                 
    Hiltner, W. A.
    1953, \apj, 118, 262
\bibitem[Hoffmeister(1929)]{hof29}                             
    Hoffmeister, C.
    1929, Astr. Nachr., 236, 235
\bibitem[Hog et al.(2000)]{tyc2}                               
    Hog, E., Fabricius, C., Makarov, V. V., Urban, S., Corbin, T.,
    Wycoff, G., Bastian, U., Schwekendiek, P. \& Wicenec, A.
    2000, \aap, 355, L27 (TYC2)
\bibitem[Keskin et al.(2000)]{kes00}                           
    Keskin, V., Yasarsoy, B. \& Sipahi, E.
    2000, Inf.\ Bull.\ Variable Stars, 4855
\bibitem[Lampens et al.(2001)]{lam01}                          
    Lampens, P., Oblak, E., Duval, D. \& Chareton, M.
    2001, \aap, 374, 132
\bibitem[Lu \& Rucinski(1999)]{ddo1}                           
    Lu, W., \& Rucinski, S. M. 
    1999, \aj, 118, 515 (Paper I)
\bibitem[Lu et al.(2001)]{ddo4}                                
    Lu, W., Rucinski, S. M. \& Ogloza, W. 
    2001, \aj, 122, 402 (Paper IV)
\bibitem[Nelson(2001)]{nel01}                                  
    Nelson, R. H.
    2001, Inf.\ Bull.\ Variable Stars, 5040
\bibitem[Nordstroem et al.(1997)]{nor97}                        
    Nordstroem, B., Stefanik, R. P., Latham, D. W., Andersen, J.
    1997, \aaps, 126, 21
\bibitem[Oezeren et al.(2001)]{oez01}                          
    Oezeren, F. F., Gunn, A. G., Doyle, J. G. \& Jevremovic, D.
    2001, \aap, 366, 202
\bibitem[Patkos \& Hempelman(1994)]{pat94}                     
    Patkos, L. \& Hempelman, A.
    1994, \aap, 292, 119
\bibitem[Pojmanski(1998)]{poj98}                               
    Pojmanski, G. 1998,
    Acta Astr., 48, 711
\bibitem[Pribulla et al.(2001)]{pri01}                         
    Pribulla, T., Vanko, M., Parimucha, S. \& Chochol, D.
    2001, Inf. Bull. Variable Stars, 5056
\bibitem[Rainger et al.(1991)]{rai91}                          
    Rainger, P. P., Hilditch, R. W. \& Edwin, R. P. 
    1991, \mnras, 248, 168
\bibitem[Rodriguez et al.(1998)]{rod98}                        
    Rodriguez, E. et al., 1998,
    \aap, 336, 920
\bibitem[Rucinski \& Duerbeck(1997)]{RD97}                
     Rucinski, S. M., \& Duerbeck, H.W. 1997, 
     \pasp, 109, 1340  (RD97)        
\bibitem[Rucinski \& Lu(1999)]{ddo2}                           
    Rucinski, S. M. \& Lu, W.
    1999, \aj, 118, 2451 (Paper II)
\bibitem[Rucinski et al.(2000)]{ddo3}                          
     Rucinski, S. M., Lu, W. \& Mochnacki, S. W.
     2000, \aj, 120, 1133 (Paper III)
\bibitem[Rucinski et al.(2001)]{ddo5}                          
     Rucinski, S. M., Lu, W., Mochnacki, S. W., 
     Ogloza, W. \& Stachowski, G.
     2001, \aj, 122, 1974 (Paper V)
\bibitem[Rucinski(2002)]{ddo7}                             
     Rucinski, S. M
     2002, \aj, submitted (Paper VII)
\bibitem[Soederhjelm(1999)]{soed99}                            
     Soederhjelm, S.
     1999, \aap, 341, 121
\bibitem[Wolf et al.(2000)]{wol00}                             
     Wolf, M., Molik, P., Hornoch, K. \& Sarounova, L.
     2000, \aaps, 147, 243
\bibitem[Yakut \& Ibanoglu(2000)]{yak00}                       
     Yakut, K. \& Ibanoglu, C.
     2000, Inf.\ Bull.\ Variable Stars, 5002

\end{thebibliography}
\end{document}


\begin{deluxetable}{lccrrrccc}

\tabletypesize{\scriptsize}

\pagestyle{empty}
\tablecolumns{9}

\tablewidth{0pt}
\tablenum{2}
\tablecaption{Spectroscopic orbital elements of the sixth set of
ten close binary systems}
\tablehead{
   \colhead{Name} &                
   \colhead{Type} &                
   \colhead{Other names} &         
   \colhead{$V_0$} &               
   \colhead{K$_1$} &               
   \colhead{$\epsilon_1$} &        
   \colhead{T$_0$ -- 2,400,000} &  
   \colhead{P (days)} &            
   \colhead{$q$}          \\       
   \colhead{}     &                
   \colhead{Sp.~type}    &         
   \colhead{}      &               
   \colhead{} &                    
   \colhead{K$_2$} &               
   \colhead{$\epsilon_2$} &        
   \colhead{(O--C)(d)~[E]} &         
   \colhead{$(M_1+M_2) \sin^3i$} &    
   \colhead{}                      
}
\startdata

SV Cam   & EA/RS & HD~44982   &  $-9.13$(0.78) & 121.86(0.76) &
  3.10 & 50,792.6597(11)  & 0.593073        & 0.641(7) \\
       & G2V     & HIP~32015  &             & 190.17(1.73) &
  7.79 & $+0.0012$~[0] & 1.871(45) &                \\[1mm]

EE Cet   & EW/W  & HD~17613   &  $+1.60$(0.93) &  84.05(1.24) &
  5.43 & 51,818.8005(7)   & 0.379917        & 0.315(5) \\
~~ADS 2163(S) & F8V & HIP~13199 &             & 266.92(1.54) &
  8.32 & $+0.0315$~[8735] & 1.706(41) &             \\[1mm]

KR Com   & EW/A  & HD~115955  & $-7.86$(0.38)  & 19.18(0.34)  &
  1.85 & 52,001.4365(4)   & 0.407968        & 0.091(2) \\
~~ADS 8863A & G0IV & HIP~65069  &             & 211.07(0.80) &
  4.45 & $+0.0430$~[8582] & 0.517(8)  &             \\[1mm]

V401 Cyg & EW/A  & BD+30$^\circ$3592 & $+25.53$(2.14) & 72.23(2.43) &
  9.74 & 51,630.9481(35)  & 0.582714        & 0.290(11) \\
         & F0V   & HIP~95816  &             & 249.13(4.53) &
  26.87 & $-0.0047$~[$-192$] & 2.008(130) &             \\[1mm] 
 
GM Dra   & EW/W  & HD~238677  & $+9.12$(1.63)     & 46.69(1.74)  &
  10.45 & 51,350.3418(13) & 0.338741        & 0.180(7)  \\
         & F5V   & HIP~84837  &             & 258.72(2.66) &
  13.69 & $-0.0058$~[$-1181$] & 1.002(42)  &             \\[1mm]

V972 Her & EW/W  & HD~164078   & $+3.98$(0.46)     & 25.95(0.36)  &
  2.34  & 51,349.1808(16) & 0.443094        & 0.167(3)  \\
         & F4V   & HIP~87958  &             & 155.65(1.00) &
  5.51  & $+0.0224$~[6430] & 0.276(6)   &             \\[1mm]

ET Leo  & EW/W   & HD~91386   & $+21.46$(0.78)      & 59.08(0.80)  &
  4.36  & 51,990.9912(13) & 0.346503        & 0.342(5)  \\
        & G8V    & HIP~51677  &             & 172.95(1.20) &
  7.30  & $+0.0021$~[10075] & 0.450(12)  &            \\[1mm]

FS Leo & EB?     & BD+15$^\circ$2335 & $-30.75$(0.64) & 53.60(0.74) &
  3.62  & 51,660.2706(14) & 0.456971        &           \\
        & F2V    & HIP~55952  &             &              &
        & $-0.0349$~[6915] &             &             \\[1mm]

V2388 Oph & EW/A & HD~163151 & $-25.88$(0.52)         & 44.62(0.48) &
  3.36  & 51,569.7076(12) & 0.802298        & 0.186(2)  \\
~~Fin~381A & F3V  & HIP~87655 &              & 240.22(0.98) &
  7.59  & $-0.0108$ [2093] & 1.926(30)   &             \\[1mm]

II UMa  & EW/A   & HD~109247 & $-8.02$(1.10)         & 43.16(0.86) &
  5.46  & 51,221.6579(44) & 0.825220        & 0.172(4)  \\
~~ADS~8594A & F5III & HIP~61237 &           & 250.96(2.74) &
  18.64 & $-0.0007$ [3298] & 2.180(80)  &             \\[1mm]

\tablenotetext{a}{SV Cam: $T_0$ has been advanced from the
published time to the time of observations.}

\tablecomments{The spectral types given in column two
are all new and relate to the combined spectral
type of all components in a system. The convention of naming the
binary components is that the subscript 1 designates
the component which is more massive so that
the mass ratio is defined to be always $q \le 1$.
The standard errors of the circular 
solutions in the table are expressed in units of last decimal places 
quoted; they are given in parantheses after each value. 
For example, the last table entry for the mass ratio 
$q$, 0.172(4), should be interpretted as $q = 0.172 \pm 0.004$.
The center-of-mass velocities ($V_0$), the 
velocity amplitudes (K$_i$) and the standard unit-weight 
errors of the solutions ($\epsilon$) are all expressed 
in km~s$^{-1}$. The spectroscopically determined 
moments of primary minima are given by $T_0$; the corresponding
(O--C) deviations (in days) have been calculated from the most 
recent available ephemerides, as given in the text,
using the assumed periods and the number of epochs given by [E].
The values of $(M_1+M_2)\sin^3i$ are in the solar mass units.
}
\enddata
\end{deluxetable}